\documentclass[%
 reprint,
 amsmath,amssymb,
 aps
]{revtex4-1}

\usepackage[utf8]{inputenc}
\usepackage{color}
\usepackage{amsmath,amssymb,amsfonts}
\usepackage{bbold}
\usepackage{color}
\usepackage{url}
\usepackage{graphicx}
\usepackage{tikz}
\usepackage{epstopdf}
\usepackage{hyperref}

\newcommand{\diag}{{\rm diag\,}}

\newcommand{\Str}{{\rm Str\,}}
\newcommand{\tr}{{\rm tr\,}}
\newcommand{\Sdet}{{\rm Sdet\,}}

\newcommand{\eins}{\leavevmode\hbox{\small1\kern-3.8pt\normalsize1}}
\newcommand{\avg}[1]{\left< #1 \right>}
\newcommand{\expleft}[1]{\exp\left[ #1 \right]}

\makeatletter
\renewcommand*\env@matrix[1][*\c@MaxMatrixCols c]{%
  \hskip -\arraycolsep
  \let\@ifnextchar\new@ifnextchar
  \array{#1}}
\makeatother

\begin{document}

\title{Eigenvalue and Eigenvector Statistics in Time Series Analysis}
\author{Paolo Barucca }
\affiliation{Department of Computer Science, University College London, London WC1E 6EA, United Kingdom}
\author{ Mario Kieburg}
\affiliation{Faculty of Physics, Bielefeld University, P.O. Box 100131, D-33501 Bielefeld, Germany}
\author{Alexander Ossipov }
\affiliation{School of Mathematical Sciences, University of Nottingham, Nottingham NG7 2RD, United Kingdom}

\date{\today}

\begin{abstract}
The study of correlated time-series is ubiquitous in statistical analysis, and the matrix decomposition of the cross-correlations between time series is a universal tool to extract the principal patterns of behavior in a wide range of complex systems. 
Despite this fact, no general result is known for the statistics of eigenvectors of the cross-correlations of correlated time-series. 
Here we use supersymmetric theory to provide novel analytical results that will serve as a benchmark for the study of correlated signals for a vast community of researchers.
\end{abstract}

\maketitle

\paragraph*{Introduction}\label{sec:intro}

The theory of complex systems ultimately deals with the identification of patterns of simple behaviours accounting for the emergence of universal dynamics in the time series measured in a vast range of disciplines, including condensed matter physics, medicine, finance, signal transmission, biology, and more recently computational social sciences~\cite{strogatz2018nonlinear}. 
A time series is a series of values scanned over time of a given observable of a system~\cite{chatfield2018introduction} such as the sea level~\cite{rahmstorf2007semi}, the temperature of a lake~\cite{sharma2015global}, the neuron activity in electroencephalography (EEG)~\cite{vseba2003random,muller2006localized}, the response in a unit of volume of a magnetic resonance imaging experiment~\cite{kwong1992dynamic}, the gross domestic product of a country~\cite{lee2005energy}, the price or return of a stock~\cite{plerou1999universal,barucca2014localization}, the volume of an order in the market~\cite{bouchaud2009markets,chiarella2009impact}, the infected individuals in a region affected by an epidemics~\cite{grenfell2001travelling}, and the online activity of a user~\cite{o2010tweets}.

The basic analysis that is ubiquitously performed when dealing with multiple time series are covariance and correlation analysis, especially with the aim of identifying the main factors accounting for time variability and parsimoniously representing the state space of the system, through denoising and dimensionality reduction. The generality of this statistical approach constitutes the basis for Principal Component Analysis (PCA)~\cite{pearson1901liii,jolliffe2011principal}, clustering analysis, and many other data mining algorithms~\cite{lloyd1982least}. In these techniques one distinguishes between eigenvalue and eigenvector statistics and both of them carry important information as we know, for example, from the theory of quantum disordered systems. Therefore it is even more surprising that only few results are available for the cross-statistics between eigenvalues and eigenvectors, when dealing with the covariance and correlation matrices of noisy time series.

The spectral density of the eigenvalues is up to now the major quantity where the theory provides robust and general results~\cite{laloux1999noise,lillo2005spectral,allez2012invariant,majumdar2012number}. For instance, the Mar\u{c}enko--Pastur distribution (MPD)~\cite{MPD} usually serves as a blueprint for describing the influence of white noise in the time series on the spectral density. Any deviation from the MPD, for instance outliers, can be considered as system specific information so that the MPD serves as a filter. However, some eigenvalues encoding relevant information might be obscured by the bulk of the spectrum described by the MPD. Then PCA may remove relevant data that should be taken into account. To distinguish those system specific eigenvalues from the eigenvalues of the MPD one needs to take into consideration the eigenvector statistics. An important step in this direction is made in the present Letter. We derive an analytical formula for the first moment of a fixed eigenvector component conditioned to a chosen eigenvalue. Moreover, we state a conjecture on their general moments and distributions for a correlation matrix of noisy time series. Our results provide insights and pave the way for a much more informative spectral decomposition in time series analysis, allowing not only to focus on the spectral density but also on the individual contribution of each component to the spectrum, leading to a much deeper understanding of a system's dynamics. 

\paragraph*{Random Matrix Model}

Specifically, we study the statistics of the eigenvectors and the eigenvalues of the matrix $C = WW^T$, with $W\in\mathbb{R}^{p\times n}$ representing $p$ time series of length $n$ or, in the case of PCA, $p$ descriptors with $n$ variants, and $W^T$ being the transpose of $W$. Thus $C$ can be interpreted as the covariance matrix between the time series aggregated in $W$ or the covariance between the descriptors respectively.
The real rectangular matrix $W$ in our model is composed of four matrices
\begin{equation}\label{Matrix-Model}
 W=\sqrt{C_L}(W_0+W_1)\sqrt{C_R},
\end{equation}
where $W_0\in\mathbb{R}^{p\times n}$ is a deterministic real matrix and $W_1\in\mathbb{R}^{p\times n}$ is a Gaussian random matrix distributed by
\begin{equation}\label{distribution}
 P(W_1)=(2\pi\sigma^2)^{-pn/2}\exp\left[-\frac{1}{2\sigma^2}\tr W_1W_1^T\right],\quad \sigma>0.
\end{equation}
The two real symmetric matrices $C_L=C_L^T\in\mathbb{R}^{p\times p}$ and $C_R=C_R^T\in\mathbb{R}^{n\times n}$ are positive definite and represent a spatio-temporal correlation between the various time series. Here, the matrix $C_L$ can be identified with a time correlation, the matrix $C_R$ with the spatial correlations, and additionally, at difference with many common models, we include an offset $W_0$. Hence $W$ is a non-centred and doubly correlated Gaussian random matrix. This form allows the model to capture in detail the case of factor models ubiquitous in statistics and econometrics. 

Though our model is quite general, it is still not the most general Gaussian random matrix model. We assume that the spatio-temporal correlations of the multivariate time series  factorize in the two matrices, $C_L$ and $C_R$. Therefore time-dependent spatial correlations, like the two epoch model~\cite{twoepoch}, are not considered here.

The random matrix model defined above can be also considered as a simple deformation of the standard real Wishart ensemble of random matrices, in which the orthogonal invariance is broken in several ways. Such non-invariant deformations of the standard random matrix ensembles were introduced and studied in different contexts including wireless communication~\cite{CD14},  vibration analysis~\cite{S03}, signal processing~\cite{NE08} and neural networks~\cite{AFM15}. There is a growing interest to the statistical properties of the eigenvectors in these ensembles. While there are some recent results about the statistics of the eigenvectors in the deformed Gaussian Orthogonal and Unitary ensembles~\cite{allez2014eigenvectors,Truong_2016,Truong_2016_EPL, BBP17,bourgade2017eigenvector,Benigni17, Truong_2018}, we are not aware of 
similar results for the Wishart ensemble except for Ref.\cite{bourgade2017eigenvector}, in which the ergodicity of the eigenvectors was proven for the special case $C_L=\eins_p$, $C_R=\eins_n$.  

In the following, we will not simply focus on the computation of the spectral density of the eigenvalues, analysed in~\cite{Recher2010,Recher2012}  with the same supersymmetric (SUSY) approach as in the present work, but also calculate a detailed eigenvector statistics of the matrix $WW^T=U\Lambda U^T$, whose eigenvalues represented by the diagonal matrix  $\Lambda=\diag(\lambda_1,\ldots,\lambda_p)$ and the eigenvectors represented by the columns of the matrix $U=\{U_{ab}\}\in\mathrm{O}(p)$. The full information about the statistics of the eigenvector components is contained in the conditional density
\begin{equation}\label{densU}
\mathcal{I}_b(\mu|\lambda)=\frac{1}{p\rho(\lambda)}{\sum}_{a=1}^p \langle\delta(\mu-|U_{ab}|^{2})\delta(\lambda-\lambda_a)\rangle,
\end{equation}
where  $b=1,\dots, p$ refers to a particular eigenvector component and
\begin{equation}\label{densrho}
\rho(\lambda)=\frac{1}{p}{\sum}_{a=1}^p \langle\delta(\lambda-\lambda_a)\rangle
\end{equation}
is the mean density of the eigenvalues and
$\langle .\rangle$ stands for the ensemble average over the distribution of $W_1$.
In the case of a factorisation of the eigenvector and eigenvalue statistics, as in the Wishart ensemble, one finds the Porter--Thomas distribution~\cite{PTD}
\begin{equation}\label{Porter}
\mathcal{I}_b^{\rm (Haar)}(\mu|\lambda)=\sqrt{\frac{p}{2\pi \mu}}\exp\left[-\frac{p\mu}{2}\right],
\end{equation} 
which is independent of the component $b$ and the eigenvalue $\lambda$ due to the Haar distributed eigenvectors. This simplification cannot be expected to hold in our non-trivial model as well as in a realistic situation. 
The computation of~\eqref{densU} or its arbitrary moments 
\begin{equation}\label{momentdef}
I_{q,b}(\lambda)=\langle \mu^q\rangle=\frac{1}{p\rho(\lambda)}\sum_{a=1}^p \langle|U_{ab}|^{2q}\delta(\lambda_a-\lambda)\rangle,
\end{equation}
where $q$ is a positive integer, is technically a very challenging problem. In this Letter we focus on the analytical derivation of the first moment $I_{1,b}(\lambda)$ and make a conjecture about an arbitrary moment $I_{q>1,b}(\lambda)$ and $\mathcal{I}_b(\mu|\lambda)$ in the conclusions.

The moments of the eigenvectors are also a standard tool to characterise properties of complex quantum systems and are used to distinguish different phases in condensed matter physics~\cite{EversMirlin2008}. Hence, we expect that it may give  valuable insights for time series as well.

Before we start with the analytical calculation of $I_{1,b}$, we want to point out that the  eigenvector components $U_{ab}$ are basis dependent. Thus the conditional distribution $\mathcal{I}_b(\mu|\lambda)$ strongly depends on the reference frame. In this work such a frame is chosen as the eigenbasis of $C_L$, allowing us to investigate the broadening of the eigenvectors due to the white noise $W_1$ and its strength $\sigma$. Another natural and valuable reference frame could be the 
eigenbasis of $\sqrt{C_L}W_0C_RW_0^T\sqrt{C_L}$ which we do not consider here for simplicity.

\paragraph*{Eigenvector Statistics with SUSY}\label{sec:evecstat}

The first moment of the eigenvectors, see~\eqref{momentdef} for $q=1$, can be computed by taking the imaginary part and the limit of a regularization $\epsilon\to 0$ of the quantity
\begin{equation}\label{momentdef.b}
I'_{1,b}(\lambda)=-\frac{1}{\sqrt{\lambda_{+}}}\left\langle \left\{\left[\begin{array}{cc} \sqrt{\lambda_{+}}\eins_p & W \\ W^T & \sqrt{\lambda_{+}}\eins_n \end{array}\right]^{-1}\right\}_{bb}\right\rangle,
\end{equation}
where $\lambda_+=\lambda+ i \varepsilon$.  Defining the $(p+n)$-dimensional unit vector $e_b$ with unity at the position $b$ and zero otherwise, this quantity can be generated by
 differentiating 
\begin{equation}\label{generate} 
Z_b(\lambda)=\left\langle \exp\left[i  \alpha^2 e_b^T\left[\begin{array}{cc} \sqrt{\lambda_{+}}\eins_p & W \\ W^T & \sqrt{\lambda_{+}}\eins_n \end{array}\right]^{-1}e_b\right]\right\rangle,
\end{equation}
with respect to the auxiliary parameter $\alpha$,  at $\alpha=0$. $\alpha$ is chosen to be real to guarantee convergence later on. Following the standard steps of the SUSY method~\cite{Recher2010,Recher2012}, we represent first the generating function $Z_b(\lambda)$ by the supersymmetric Gaussian integral, average over the random matrix $W_1$ and finally apply the Hubbard-Stratonovich transformation~\cite{fyodorov2008hyperbolic}. In this way, we derive the following supersymmetric representation for $I_{1,b}(\lambda)$ (see the Supplemental Material~\cite{supplmat} for details),
\begin{equation}\label{I_1-G-rep}
I_{1,b}(\lambda)= \frac{\int d[T]\expleft{F(T_+,T_-)}G_{1b,1b}\sqrt{\Sdet G}}
{\sqrt{\lambda_+}\int d[T]\expleft{F(T_+,T_-)}},   
\end{equation}
where $F(T_+,T_-)=-\frac{1}{2\sigma^2}\Str(T_+^2+T_-^2)-\Str T_+L$, $L=\diag(-\eins_2;\eins_2)$. The  $(2|2)\times(2|2)$ supermatrices $T_{\pm}$ are symmetric in the boson-boson block and self-dual in the fermion-fermion block and their eigenvalues run along complex contours that are detailed in the Supplemental Material~\cite{supplmat}. The supersymmetric Green function $G$ has the form
\begin{widetext}
\begin{equation}\label{Gdef}
G=\left[\begin{array}{cc} \sqrt{\lambda_{+}}\eins_p\otimes LJ-C_L\otimes(T_+-i  T_-+\sigma^2L)LJ & \sqrt{C_L}W_0\sqrt{C_R}\otimes LJ \\ \sqrt{C_R}W_0^T\sqrt{C_L}\otimes LJ & \sqrt{\lambda_{+}}\eins_{n}\otimes LJ-C_R\otimes(T_++i  T_-+\sigma^2L)LJ \end{array}\right]^{-1},
\end{equation}
\end{widetext}
with $J=\diag(\eins_{2} ; \tau_2)$.
The representation~\eqref{I_1-G-rep} is exact, but rather involved and technical. An expression for the mean level density can be obtained by summing over $b=1,\ldots,p$ and should be compared with the corresponding result in~\cite{Recher2010,Recher2012,waltner2015eigenvalue}. The above expression simplifies a lot in the limit $n,p\to \infty$, which is considered next.

\begin{figure}[t!]
\includegraphics[width=86mm]{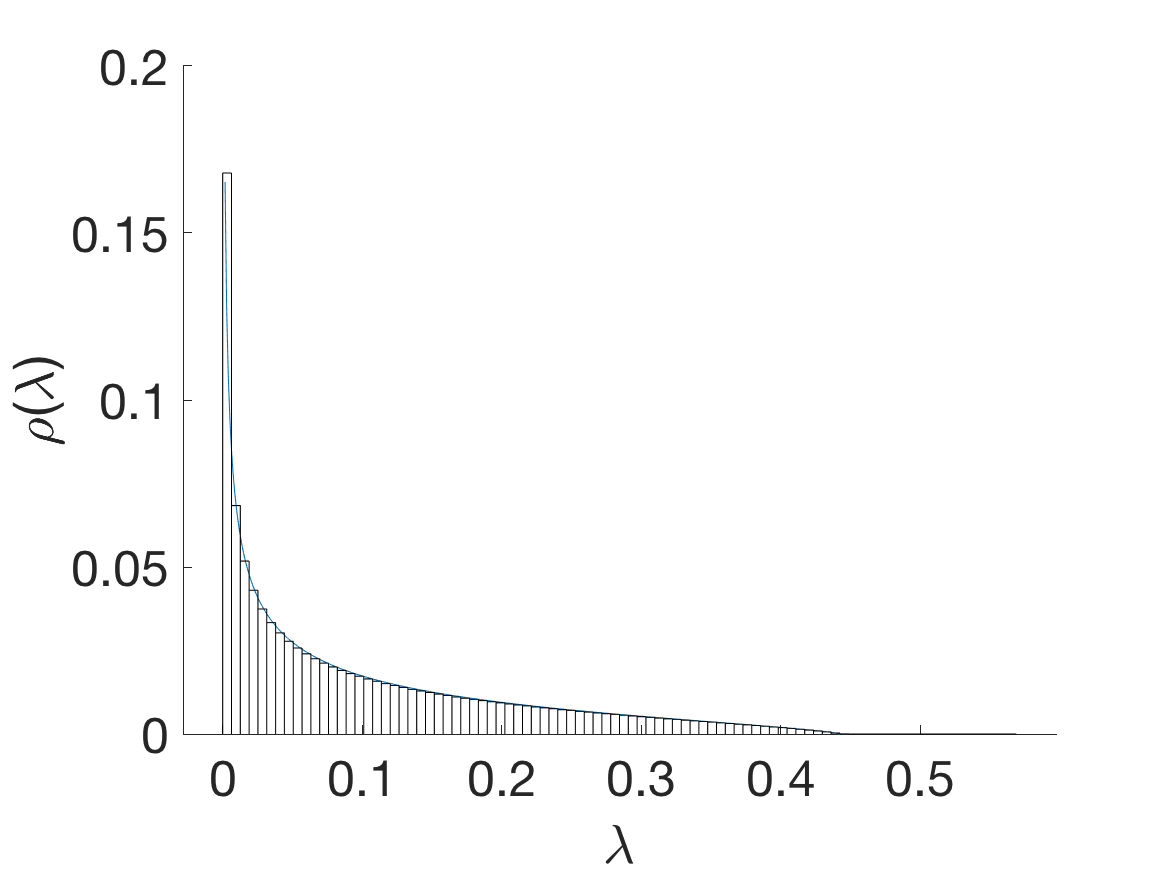}
\caption{Eigenvalue density for the one-factor model: analytical result (solid line, combination of Eqs.~\eqref{eq:leveldens},\eqref{sol.a}, and~\eqref{sol.b}) and Monte-Carlo simulation (histogram, $p=2000$, $n=2000$ and sample size is $1000$). $(C_{L})_{ij} = l_i^{-1}\delta_{ij}$, $C_R = \eins_n$,  $(W_0)_{it} = w_ix_t$, $\sigma=0.3/\sqrt{n}$, where the $l_i$s and $w_i$s are drawn only once from a log-normal distribution with mean $1$ and variance $e^{\sigma_{L/w}^2}-1$, with $\sigma_L=0.3$ and $\sigma_w=0.1$, respectively, and then kept fixed. The vector $\{x_t=A \cos(f\,t)\}$ is a cosine wave with frequency $f=1/50$ and amplitude $A=1/\sqrt{n}$.}
\label{fig:exp1}
\end{figure}

\paragraph*{Macroscopic level density and limiting statistics}

In most applications, one is interested in the limit $n,p\to \infty$.  In this limit the integral in Eq.\eqref{I_1-G-rep} can be evaluated using the saddle-point approximation. 
To derive the saddle-point equation, it is convenient to introduce the supermatrices $S=T_+-i  T_-+\sigma^2L$ and $R=T_++i  T_-+\sigma^2L$, which can be considered as independent. The saddle-point solution contributing most to the integral is given by the diagonal matrices $S_0=s_0\eins_{2|2}$ and $R_0=r_0\eins_{2|2}$ with the complex parameters $s_0$ and $r_0$ that satisfy the coupled  equations~\cite{supplmat}
\begin{equation}\label{saddle.eq.scalar}
\begin{split}
&\frac{r_0}{\sigma^2}=\tr\left[A_L-W_0A_R^{-1}W_0^T \right]^{-1},\\
 &\frac{s_0}{\sigma^2}=\tr\left[A_R-W_0^TA_L^{-1}W_0 \right]^{-1},\ {\rm with}\\
&A_L=\sqrt{\lambda}C_L^{-1}-s_0\eins_p\ {\rm and}\ A_R=\sqrt{\lambda}C_R^{-1}-r_0\eins_n.
\end{split}
\end{equation}
The mean level density is up to a normalisation constant given by
\begin{equation}\label{eq:leveldens}
\begin{split}
 \rho(\lambda)\propto & {\rm Im}\left[\tr\left( Q^{-1}\diag(\eins_p,0)\right)\right],\\
 Q=&
\left[\begin{array}{cc} \sqrt{\lambda}\eins_p-s_0C_L & \sqrt{C_L}W_0\sqrt{C_R} \\ \sqrt{C_R}W_0^T\sqrt{C_L} & \sqrt{\lambda}\eins_n-r_0C_R \end{array}\right],  
\end{split}
\end{equation}
where we assume $p\leq n$ without loss of generality. The case $p>n$ only yields an additional Dirac delta function at the origin. The formula~\eqref{eq:leveldens} reduces to the MPD~\cite{MPD} in the case of the Wishart ensemble, i.e., $C_L=\eins_p$, $C_R=\eins_n$ and $W_0=0$. We illustrate the result for $\rho(\lambda)$ in Fig.~\ref{fig:exp1} for the one-factor model, which is described in the next subsection.

The result for $I_{1,b}(\lambda)$ can be expressed in terms of the same matrix $Q$ and reads
\begin{equation}\label{eq:expr_Iqb}
I_{1,b}(\lambda)=\frac{{\rm Im}\left[\tr \left(Q^{-1}\diag(\widehat{E}_b,0)\right)\right]}
{{\rm Im}\left[\tr\left(Q^{-1}\diag(\eins_p,0)\right)\right]},
\end{equation}
which constitutes the main result of the present Letter. The normalisation is fixed by the condition $\sum_{b=1}^pI_{1,b}(\lambda)=1$. We note that for a Haar distributed vector one has $I_{1,b}^{\rm (Haar)}(\lambda)=1/p$.

\begin{figure}[t!]
\includegraphics[width=86mm]{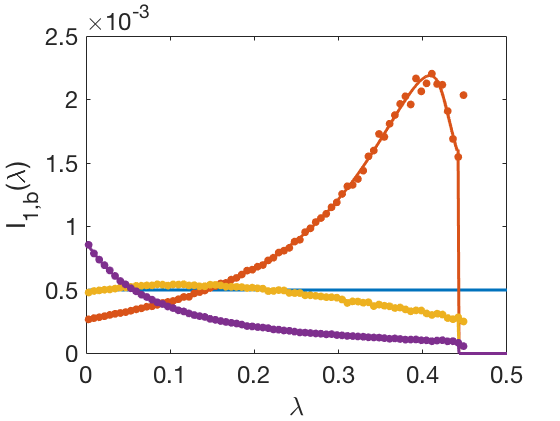}
\caption{Moments $I_{1,b}(\lambda)$ of the eigenvectors (\ref{eq:expr_Iqb}) for the one-factor model corresponding to different components: analytical result (solid line) and Monte Carlo simulation (points),  for the same parameters as in Fig.~\ref{fig:exp1}. Moments of the components corresponding to the 50th (orange, $(1/l)_{50} = 1.88$), 1000th (yellow, $(1/l)_{1000} = 1.044$), and 1950-th (purple, $(1/l)_{1950} = 0.57$) values of $1/l$. The blue line corresponds to $I_{1,b}^{\rm (Haar)}(\lambda)=1/p$. }
\label{fig:exp2}
\end{figure}

\paragraph*{One-factor model}

To illustrate our findings we apply our general results to the one-factor model supplemented with Gaussian noise. Specifically,  we set $ W_{0} = wx^T$, where $w$ and $x$ are column vectors of length $p$ and $n$, respectively. The correlation matrices are chosen to be diagonal $C_{L} = \diag(l_1^{-1},\ldots,l_p^{-1})$ and $ C_{R}=\eins_n $.
The vector $x$ represents a common factor, e.g. the market mode in financial time series analysis, and the component $w_j$ quantifies the relative weight of the common factor on the $j$th time series, before normalization. 

\begin{figure}[t!]
\includegraphics[width=86mm]{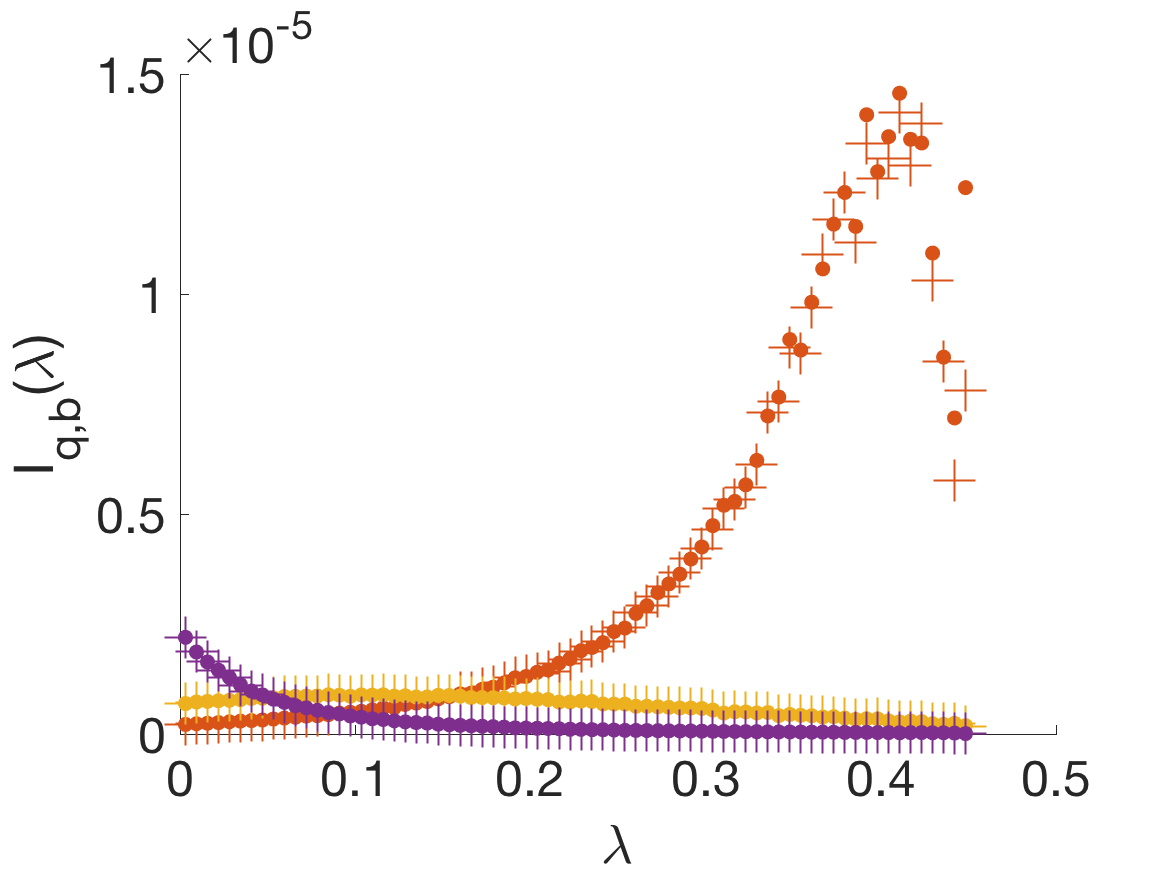}
\caption{Moments $I_{2,b}(\lambda)$ (dots) of the eigenvectors for the one-factor model and  $\frac{(2q)!}{2^q\, q!}[I_{1,b}(\lambda)]^q$ (Eq.\eqref{conj1}) for q=2 (crosses) as functions of $\lambda$, for the same parameters as in Fig.~\ref{fig:exp2}.}
\label{fig:exp3}
\end{figure}

We plug the matrices of the one-factor model into the saddle-point equation~\eqref{saddle.eq.scalar} and simplify the resulting expression via the Sherman-Morrison identity for the inverse matrices~\cite{sherman1950adjustment}, i.e. $(A + uv^T)^{-1} = A^{-1} - \frac{A^{-1}uv^TA^{-1}}{1 + v^TA^{-1}u} $. This leads to the coupled equations 
\begin{equation}\label{sol.a}
\begin{split}
\frac{r_0}{\sigma^2} =& \sum_j \frac{1}{\sqrt{\lambda}l_j - s_0} \\
&+ \frac{\sum_tx_t^2\sum_jw_j^2/(\sqrt{\lambda}l_j - s_0)^2}{ \sqrt{\lambda}-r_0- \sum_tx_t^2\sum_j w_j^2/(\sqrt{\lambda}l_j-s_0)},\\
\frac{s_0}{\sigma^2} =&\frac{1}{\sqrt{\lambda} - r_0}\biggl(n \\
&+\frac{\sum_tx_t^2\sum_jw_j^2/(\sqrt{\lambda}l_j - s_0)}{ \sqrt{\lambda}-r_0- \sum_tx_t^2\sum_j w_j^2/(\sqrt{\lambda}l_j-s_0)}\biggl).
\end{split}
\end{equation}
Solving these saddle-point equations we can derive the spectral density and the moments of the eigenvectors,  simply by plugging the following matrix elements in Eqs.~\eqref{eq:leveldens}-\eqref{eq:expr_Iqb}
\begin{equation}\label{sol.b}
\begin{split}
(Q^{-1})_{ii} = & \frac{l_i}{\sqrt{\lambda}l_i - s_0} +\frac{1}{(\sqrt{\lambda}l_i-s_0)^2} \\
&\times  \frac{l_iw_i^2\sum_tx_t^2}{\sqrt{\lambda}-r_0-\sum_tx_t^2\sum_j w_j^2 /(\sqrt{\lambda}l_j - s_0)}.
\end{split}
\end{equation}
We illustrate these results in Figs.~\ref{fig:exp1} and~\ref{fig:exp2}, where we also compare them with Monte-Carlo simulations. The deviations from the Porter-Thomas distribution~\eqref{Porter}, which  yields for the first moment the constant $I_{1,b}^{\rm (Haar)}=1/p$, can be readily seen for some components of the eigenvectors. They indicate that the corresponding eigenvalues still carry a lot of information on the matrix $C_L$, although these eigenvalues are evidently inside the bulk of the spectrum, cf., Fig~\ref{fig:exp1}. This simple example demonstrates the strength of the combined statistics of eigenvalues and eigenvectors.

\paragraph*{Conclusions}

The general result in Eq.~\eqref{eq:expr_Iqb} provides a powerful analytical methodology to quantify the expected value of the square of specific components in a given eigenvalue interval for a wide range of random matrices. 
We tested numerically these analytical results in detail for the one-factor model (see Figs.~\ref{fig:exp1}-\ref{fig:exp2}).
Our general formulation allows an arbitrary number of factors to be added in the matrix $W_0$. Although our analytical results were derived in the limit $n,\:p \to \infty$, they show a very good agreement with the results of numerical simulations at finite $n$ and $p$.
The rate of convergence to the limiting statistics will generally depend on the input $W_0$, $C_L$, and $C_R$.

In the present work we derived analytically a closed result only for the first moment  $I_{1,b}(\lambda)$ of an eigenvector under the condition of a fixed eigenvalue. However we conjecture that all higher moments are related to the first moment as follows:
\begin{equation}\label{conj1}
I_{q,b}(\lambda) =\frac{(2q)!}{2^q\, q!}[I_{1,b}(\lambda)]^q,
\end{equation}
which corresponds to a locally rescaled Porter-Thomas distribution
\begin{equation}\label{conj}
\mathcal{I}_b(\mu|\lambda)=\frac{1}{\sqrt{2\pi I_{1,b}(\lambda)\mu}}\exp\left[-\frac{\mu}{2I_{1,b}(\lambda)}\right].
\end{equation} 
A similar result has been also found for the conditioned eigenvector statistics of the deformed Gaussian Unitary Ensemble (GUE) in~\cite{Truong_2016,Truong_2016_EPL}. The only difference is the prefactor in~\eqref{conj1}, which is equal to  $(2q)!/(2^q\, q!)$ in our case and given by  $q!$ for the complex eigenvectors in the deformed GUE \cite{Truong_2016,Truong_2016_EPL}. These numerical values result from the averaged moments of real and complex normalized vectors, respectively. We have tested this conjecture numerically for $q=2$ and found a nice agreement, see  Fig.~\ref{fig:exp3}. 

We are confident that our analytical results are of general relevance for the spectral decomposition of time series and could lead to unprecedented understanding of the full statistics of the eigen-components in signal analysis. A strong deviation of the moment $I_{q,b}(\lambda)$ from the constant $(2q)!/(2^q\, q!)$ hints at an eigenvector-eigenvalue pair that contains system specific information. This knowledge can improve PCA and other techniques to reduce highly dimensional data without loosing relevant information.

\acknowledgements

MK acknowledges financial support by  the  German  research
council  (DFG)  through  CRC  1283:  ``Taming uncertainty  and  profiting  from  randomness
and low regularity in analysis, stochastics and their applications''.
PB acknowledges support from the London Institute for Mathematical Sciences (LIMS).

\bibliography{Wishart-v2}
\bibliographystyle{apsrev4-1}


\widetext
\clearpage
\begin{center}
\textbf{\large Supplemental Material: Eigenvalue and Eigenvector Statistics in Time Series Analysis}
\end{center}
\setcounter{equation}{0}
\setcounter{figure}{0}
\setcounter{table}{0}
\setcounter{section}{0}
\setcounter{page}{1}
\makeatletter
\renewcommand{\theequation}{S\arabic{equation}}

\section{Derivation of the supersymmetric integral representation for the moments of the eigenvectors}
The quantity $I'_{1,b}$ defined in Eq.\eqref{momentdef.b} can be computed by differentiating the generating function \eqref{generate}
\begin{equation}\label{supple-generate}
Z_b(\lambda)=\left\langle \exp\left[i  \alpha^2 e_b^T\left[\begin{array}{cc} \sqrt{\lambda_{+}}\eins_p & W \\ W^T & \sqrt{\lambda_{+}}\eins_n \end{array}\right]^{-1}e_b\right]\right\rangle
\end{equation}
with respect to $i\alpha^2$ and setting $\alpha=0$. The normalization is given as $\lim_{\lambda\to\infty}Z_b(\lambda)=1$. In order to construct a representation of $Z_b(\lambda)$ in terms of the supersymmetric integral we use the identity
\begin{equation}\label{Gausstrick}
\begin{split}
&\exp\left[i \alpha^2 e_b^T\left[\begin{array}{cc} \sqrt{\lambda_{+}}\eins_p & W \\ W^T & \sqrt{\lambda_{+}}\eins_n \end{array}\right]^{-1}e_b\right]\int d[\psi]d[\phi]\exp\left[-\Str(\phi,\psi)^T(\phi,\psi)J\right]\\
=&\int d[\psi]d[\phi]\exp\left[i  \Str(\phi,\psi)^T\left[\begin{array}{cc} \sqrt{\lambda_{+}}\eins_p & W \\ W^T & \sqrt{\lambda_{+}}\eins_n \end{array}\right](\phi,\psi)J+2\alpha\Str(\phi,\psi)^T(e_b,0,0,0)\right],
\end{split}
\end{equation}
where we employed the matrix $\psi$ which is an $(n+p)\times 2$ dimensional matrix of real Grassmann variables and $\phi$ is an $(n+p)\times2$ dimensional ordinary real matrix. The two matrices are introduced in order to cancel the resulting determinants from the Gaussian integral. To ensure integrability we have introduced the constant matrix $J=\diag(\eins_{2} ; \tau_2)$, where $\tau_2$ is the second Pauli matrix. 

To simplify the notation, we define the diagonal $(2|2)\times(2|2)$ supermatrix $L=\diag(-\eins_2;\eins_2)$ and the $(n+p)\times(2|2)$ rectangular supermatrix $E_b=(\alpha e_b,0;0,0)$. Moreover, we rearrange the matrices $\psi$ and $\phi$ in the $p\times(2|2)$ supermatrix $V_L$ and the $n\times(2|2)$ supermatrix $V_R$ as follows
\begin{equation}
(\phi,\psi)=\left(\begin{array}{c} V_L \\ V_R \end{array}\right).
\end{equation}
Both matrices are two real rectangular supermatrices $V_L=V_L^*$ and $V_R=V_R^*$ with dimensions $p\times(2|2)$ and $n\times(2|2)$ respectively. The first two columns of $V_L$ and $V_R$ are real variables while the last two columns are Grassmann variables. In this way, we find
\begin{equation}
\begin{split}
Z_b(\lambda)&=\frac{\avg{\int d[V_R,V_L]\expleft{S(V_R,V_L,W)}}}{\int d[V_R,V_L]\expleft{S_0(V_R,V_L)}},\\
S(V_R,V_L,W)&=-i \sqrt{\lambda_{+}}\Str LJ (V_L^TV_L+V_R^TV_R)-i \Str L J(V_L^TWV_R+V_R^TW^TV_L)+2\Str(V_L,V_R)^TE_b,\\
S_0(V_R,V_L)&=-\Str J(V_L^TV_L+V_R^TV_R).
\end{split}
\end{equation}

The average over $W_1$ yields
\begin{equation}
\begin{split}
Z_b(\lambda)=&\frac{\int d[V_R,V_L]\expleft{S_1(V_R,V_L)+S_2(V_R,V_L)}}{\int d[V_R,V_L]\expleft{S_0(V_R,V_L)}},\\
S_1(V_R,V_L)=&-i \sqrt{\lambda_{+}}\Str LJ(V_L^TV_L+V_R^TV_R)-i \Str L J(V_L^T\sqrt{C_L}W_0\sqrt{C_R}V_R+V_R^T\sqrt{C_R}W_0^T\sqrt{C_L}V_L)\\
&+2\Str(V_L,V_R)^TE_b,\\
S_2(V_R,V_L)=&-2\sigma^2\Str V_L^TC_LV_LLJV_R^TC_RV_RLJ.
\end{split}
\end{equation}
Since the action contains a quartic term in the matrices $V_L$ and $V_R$, the next step is to perform the Hubbard-Stratonovich transformation, which allows one to decouple such terms.  
Up to the normalization the result reads
\begin{equation}
\begin{split}\label{SUSY.1}
Z_b(\lambda)\propto&\int d[V_R,V_L]\int d[T]\exp\left[S_1(V_R,V_L)+S_3(V_R,V_L, T_+,T_-)\right],\\
S_3(V_R,V_L, T_+,T_-)=&i \sigma^2\Str J(V_L^TC_LV_L +V_R^TC_RV_R)-\frac{1}{2\sigma^2}\Str(T_+^2+T_-^2)\\
&+i \Str T_+JL(V_L^TC_LV_L +V_R^TC_RV_R+i  J)+\Str T_-JL(V_L^TC_LV_L -V_R^TC_RV_R).
\end{split}
\end{equation}
The parametrization of the two $(2|2)\times(2|2)$ supermatrices $T_{\pm}$ needs to be chosen carefully to guarantee the convergence of the integral. They are given by
\begin{equation}\label{auxmatrix}
T_+=\left[\begin{array}{cc} B_1+i  C(B_2) & \eta_1\hat\tau_2 \\ -\eta_1^T & i  F_1 \end{array}\right]\ {\rm and}\ T_-=\left[\begin{array}{cc} B_2 & \eta_2\hat\tau_2 \\ -\eta_2^T & i  F_2 \end{array}\right]
\end{equation}
equipped with the flat Berezinian measure
\begin{equation}
d[T]=d[B_1]d[B_2]d[F_1]d[F_2]d[\eta_1]d[\eta_2].
\end{equation}
The ordinary matrices $B_1$ and $B_2$ are negative definite and symmetric and can be diagonalized with orthogonal matrices $O_1,O_2\in{\rm O}(2)$
as follows
\begin{equation}\label{para.a}
B_1=-O_1b_1O_1^{-1}\ {\rm and}\ B_2=-O_2b_2O_2^{-1},
\end{equation}
with $b_1,b_2$ two positive definite diagonal matrices. The matrix $C(B_2)$ has the form
\begin{equation}\label{para.b}
C(B_2)=-O_2\sqrt{\eins_{2}+b_2^2}O_2^{-1}.
\end{equation}
 The matrices $F_1$ and $F_2$ are Hermitian self-dual matrices and $\eta_1$ and $\eta_2$ are two $2\times 2$ rectangular matrices  whose entries are independent real Grassmann variables.

The shift of $B_1$ in $T_+$ by the imaginary part $\sqrt{\eins_{2}+B_2^2} $ solves a convergence problem in the Gaussian terms in~\eqref{SUSY.1}. In particular the Gaussian integrals over the supermatrices $V_L$ and $V_R$ are absolutely convergent and yield
\begin{equation}
Z_b(\lambda) \propto\int d[T]\exp\left[-\frac{1}{2\sigma^2}\Str(T_+^2+T_-^2)-\Str T_+L-i \alpha^2G_{1b,1b}\right]\sqrt{\Sdet G}, 
\end{equation}
where $G$ is defined as in~\eqref{Gdef}.
Hence, $G_{\mu a,\nu b}$ has four indices with $\mu,\nu=1,\ldots,4$ and $a,b=1,\ldots,n+p$. To fix the normalization we take $\lambda\to\infty$ and notice that $G$ becomes approximately $\lambda_{+}^{-1/2}\eins_{n+p}\otimes LJ$. Therefore we end up with the intermediate result
\begin{equation}\label{part.inter.1}
\begin{split}
Z_b(\lambda)=&\frac{\int d[T]\exp\left[-\frac{1}{2\sigma^2}\Str(T_+^2+T_-^2)-\Str T_+L-i \alpha^2G_{1b,1b}\right]\sqrt{\Sdet G}}{\int d[T]\exp\left[-\frac{1}{2\sigma^2}\Str(T_+^2+T_-^2)-\Str T_+L\right]}.
\end{split}
\end{equation}

Coming back to our original problem we notice that we are interested in the first derivative with respect to $i \alpha^2$ at $\alpha=0$. In particular, the quantity $I_{1,b}(\lambda)$ is given by
\begin{equation}\label{part.inter.2}
I_{1,b}(\lambda)=\frac{1}{\sqrt{\lambda_{+}}}\frac{\int d[T]\exp\left[-\frac{1}{2\sigma^2}\Str(T_+^2+T_-^2)-\Str T_+L\right]G_{1b,1b}\sqrt{\Sdet G}}{\int d[T]\exp\left[-\frac{1}{2\sigma^2}\Str(T_+^2+T_-^2)-\Str T_+L\right]},
\end{equation}
which coincides with Eq.\eqref{I_1-G-rep}.

\section{Saddle-point equation}

For deriving the saddle-point equation we only need to consider the exponential function and the superdeterminant in the integral~\eqref{part.inter.2}. The term $G_{1b,1b}$ is only a polynomial prefactor which does not influence the saddle-point solution.
It is easier to study the saddle-point by introducing the supermatrices $S=T_+-i  T_-+\sigma^2L$ and $R=T_++i  T_-+\sigma^2L$, which can be considered to be independent. Then the action, i.e. the function that need to be minimised, is
\begin{equation}
\frac{1}{2\sigma^2(n+p)}\Str SR+\frac{1}{2(n+p)}\Str {\rm ln}\left[\begin{array}{cc} \sqrt{\lambda_{+}}  C_L^{-1}\otimes\eins_{2|2}- \eins_p\otimes S &   W_0\otimes \eins_{2|2} \\   W_0^T\otimes \eins_{2|2} & \sqrt{\lambda_{+}}  C_R^{-1}\otimes \eins_{2|2}- \eins_n\otimes R \end{array}\right].
\end{equation}
Differentiating it with respect to $S$ and $R$ yields two coupled equations
\begin{equation}\label{saddle.eq}
\begin{split}
R-\sigma^2\tr_1\left[\sqrt{\lambda_{+}}C_L^{-1}\otimes\eins_{2|2}-(W_0\otimes\eins_{2|2})(\sqrt{\lambda_{+}}C_R^{-1}\otimes \eins_{2|2}-\eins_n\otimes R)^{-1}(W_0^T\otimes\eins_{2|2}) -\eins_p\otimes S\right]^{-1}=&0,\\
S-\sigma^2\tr_1\left[\sqrt{\lambda_{+}}C_R^{-1}\otimes\eins_{2|2}-(W_0^T\otimes\eins_{2|2})(\sqrt{\lambda_{+}}C_L^{-1}\otimes \eins_{2|2}-\eins_p\otimes S)^{-1}(W_0\otimes\eins_{2|2}) -\eins_n\otimes R\right]^{-1}=&0.
\end{split}
\end{equation}
The operator $\tr_1$ is the partial trace over the first tensor space which is here the space of ordinary $n\times n$ and $p\times p$ matrices, respectively. 

The saddle-point equation is rotation invariant, i.e., when $(S_0,R_0)$ is a solution then this is also true for $(R_0S_0R_0^{-1},R_0)$ as well as $(S_0,S_0R_0S_0^{-1})$ and any kind of combination. This can be seen by multiplying both equations from the left and the right with $R$ and $R^{-1}$, which is equivalent to replacing $S$ by $RSR^{-1}$. Assuming that the saddle-point solution $(S_0,R_0)$ is unique, we conclude then that $S_0$ and $R_0$ must commute. The uniqueness of the solution should follow from the fact the contour of integration, which was shifted by the term $i\epsilon$, can't cross the poles and the fact that the Berezinian (the Jacobian in superspace), that is  $|b_{1j}-b_{2j}|/[(b_{1j}-if_j)^2(b_{1j}-if_j)^2]$ for $j=1,2$, is not suppressed only when the multiplicity of the eigenvalues in the Fermion-Fermion blocks is equal to those in the Boson-Boson block.  The Fermion-Fermion blocks are doubly degenerate due to their Hermitian self-duality. Thus also the Boson-Boson blocks are doubly degenerate, which implies for $(2|2)\times (2|2)$ supermatrices that we can diagonalize $S$ and $R$ simultaneously and the solution has to be diagonal and degenerate, i.e., $S_0=s_0\eins_{2|2}$ and $R_0=r_0\eins_{2|2}$. Substituting this ansatz into Eq.~\eqref{saddle.eq} we derive Eq.\eqref{saddle.eq.scalar},  which is
\begin{equation}\label{suppl-saddle.eq.scalar}
\begin{split}
\frac{r_0}{\sigma^2}=\tr\left[\sqrt{\lambda_{+}}C_L^{-1}-W_0(\sqrt{\lambda_{-}}C_R^{-1}-r_0\eins_n)^{-1}W_0^T -s_0\eins_p\right]^{-1},\\
\frac{s_0}{\sigma^2}=\tr\left[\sqrt{\lambda_{+}}C_R^{-1}-W_0^T(\sqrt{\lambda_{-}}C_L^{-1}-s_0\eins_p)^{-1}W_0 -r_0\eins_n\right]^{-1}.
\end{split}
\end{equation}
The $\epsilon$ regularization only determines which saddle-point has to be chosen, especially which sign the imaginary part carries. Assuming the correct sign of the imaginary part we neglected this regularization in Eq.~\eqref{saddle.eq.scalar}. 
\end{document}